# Competing Orbital Magnetism and Superconductivity in electrostatically defined Josephson Junctions of Alternating Twisted Trilayer Graphene

Vishal Bhardwaj[1†], Lekshmi Rajagopal[1†], Lorenzo Arici[1], Matan Bocarsly[1], Alexey Ilin[1], Gal Shavit[5], Kenji Watanabe[3], Takashi Taniguchi[4], Yuval Oreg[1], Tobias Holder[6], Yuval Ronen[1*]

**Abstract:**

The coexistence of superconductivity and magnetism within a single material system represents a long-standing goal in condensed matter physics. Van der Waals-based moiré superlattices provide an exceptional platform for exploring competing and coexisting broken symmetry states. Alternating twisted trilayer graphene (TTG) exhibits robust superconductivity at the magic angle of 1.57° and 1.3°, with suppression at intermediate twist angles. In this study, we investigate the intermediate regime and uncover evidence of orbital magnetism. As previously reported, superconductivity is suppressed near the charge neutrality point (CNP) and emerges at larger moiré fillings. Conversely, we find orbital magnetism most substantial near the CNP, diminishing as superconductivity develops. This complementary behavior is similarly observed in the displacement field phase space, highlighting a competitive interplay between the two phases. Utilizing gate-defined Josephson junctions, we probe orbital magnetism by electrostatically tuning the weak links into the magnetic phase, revealing an asymmetric Fraunhofer interference pattern. The estimated orbital ferromagnetic ordering temperature is approximately half the superconducting critical temperature, coinciding with the onset of Fraunhofer asymmetry. Our findings suggest that the observed orbital magnetism is driven by valley polarization and is distinct from the anomalous Hall effect reported at integer fillings in twisted graphene systems. These results offer insights into the interplay between superconductivity and magnetism in moiré superlattices.

───────────────────────────────

[1]Department of Condensed Matter Physics, Weizmann Institute of Science, Rehovot 7610001, Israel

[3]Research Center for Functional Materials, National Institute for Materials Science, Tsukuba 305-0044, Japan

[4]International Center for Materials Nanoarchitectonics, National Institute for Materials Science, Tsukuba 305-0044, Japan

[5]Department of Physics and Institute for Quantum Information and Matter, California Institute of Technology, Pasadena, California 91125, USA

[6]School of Physics and Astronomy, Tel Aviv University, Tel Aviv 69978, Israel

[†]These authors contributed equally to this work
[*]yuval.ronen@weizmann.ac.il

Exploring exotic phases arising from strong electron-electron interactions has captivated the condensed matter physics community, unveiling a variety of correlated phases such as superconductivity, magnetism, and topological states[1,2]. Flat energy bands, where the kinetic energy is suppressed, have emerged as a fertile ground for these phases, as the dominance of interaction energy facilitates the emergence of correlated phenomena[3–5]. Twisted van der Waals (vdW) heterostructures, with their tunable moiré superlattices, have led to groundbreaking discoveries, including superconductivity [6–8], orbital magnetism [9,10], Chern insulators[11,12], and, more recently, zero magnetic field fractionally correlated insulators[13]. Among these phenomena, orbital magnetism in twisted graphene systems stands out due to its potential to coexist with superconductivity, opening avenues for realizing non-abelian anyons in hybrid systems—a highly sought-after goal in condensed matter physics[14]. However, understanding the interplay between these two phases remains a significant challenge.

Twisted bilayer graphene (TBG) at the 'magic angle' of ~1.1° was the first system to demonstrate flat bands and correlated insulating and superconducting phases[6,15,16]. The twist angle between the atomically thin layers creates a moiré superlattice, where electrons can become asymptotically localized. This localization results from the fine-tuned destructive interference of wavefunction overlaps at the atomic scale, leading to a minimal net kinetic energy. Introducing a third graphene layer, with the top and bottom layers mirror-symmetric relative to the middle layer ($\theta_{TM} = -\theta_{MB} = 1.57°$), creates a moiré superlattice with a periodicity of approximately 9 nm. At this angle, the twisted trilayer graphene (TTG) band structure develops flat bands and an additional gapless Dirac cone characteristic of monolayer graphene (MLG) [7,17]. TTG offers a tunable platform to study superconductivity, which depends on density and displacement field. It has shown robust superconductivity compared to TBG, persisting over a wide range of twist angles from 1.57° to approximately 1.2°, and enduring magnetic fields that violate the Pauli limit [18,19], hinting towards an unconventional nature. Surprisingly, superconductivity is suppressed at intermediate twist angles around 1.4° [19,20].

Meanwhile, orbital magnetism has been observed in graphene-based heterostructures with broken inversion symmetry due to alignment with hexagonal boron nitride (hBN)[21–23] and in off-magic-angle unaligned TBG [24]. However, these systems typically lack the coexistence of superconductivity and orbital magnetism, as time-reversal symmetry breaking, required for magnetism, tends to destabilize superconducting phases. Theoretically, superconductivity and orbital magnetism can emerge in flat-band systems driven by strong interactions. The coexistence of these phases has been reported in TBG, where superconductivity accompanies anomalous Hall effect at a moiré filling factor of ν = 1[25]. Yet, the competition and coexistence of these phases remain poorly understood in TTG, particularly in intermediate twist angles.

In this work, we report the competition of superconductivity and OM in alternating TTG heterostructures with three previously unexplored twist angles of 1.38°, 1.41° and 1.44°. We observe OM through sharp slope changes (jumps) in the Hall resistance around zero out-of-plane magnetic fields near the charge neutrality point in both electron and hole doping regions, accompanied by weak hysteresis. OM persists up to moiré densities or displacement fields where superconductivity emerges; see Fig. 1a for a schematic phase diagram of OM and superconductivity in TTG as a function of twist angle, $\theta$, moiré filling factor, $\nu$, and displacement field, $D$. We employed an electrostatically defined Josephson junction to probe the nature of the weak link, tuning it between different phases. When the weak link is tuned to the OM phase, we observe an asymmetric Fraunhofer pattern, likely caused by current-induced magnetization switching. The ordering temperature of the magnetic phase is estimated to be ~650 mK, less than the critical temperature for superconductivity (1.3K). These observations suggest that, in this twist angle range, TTG exhibits key differences in its phase diagram compared to twisted bilayer graphene (TBG), with a distinct interplay between superconducting and magnetic phases.

Our vdW heterostructures consist of alternating TTG layers, with the bottom and top graphene layers twisted mirror-symmetrically relative to the middle layer. The TTG heterostructure is encapsulated by top and bottom insulating hBN layers with a 25 nm and 30 nm thickness, respectively, and placed on top of a Si/SiO2 substrate serving as a back gate. We fabricate the heterostructure into three consecutively connected Hall bars, A, B, and C, by introducing a Ti/Au top gate. We make use of the variation in the angle along the TTG heterostructure to study the OM (Fig. 1b) and make use of the linked regions between A-B (200 nm) and B-C (100 nm) to form Josephson junctions to study supercurrent flow-mediated *via* the OM region (Fig. 1b). For additional information on the heterostructure and device fabrication, see supplementary information (SI) section 1 and Fig. S1.

The carrier density, $n$, and displacement field, $D$, in the TTG are electrostatically tuned by both top and bottom gate voltages ($V_{BG}$, $V_{TG}$) coupled by their respective capacitances ($C_{BG}$, $C_{TG}$), see SI section 1. The device transport properties were measured in a dilution refrigerator with a base temperature of 10 mK equipped with a 9-1-1 T vector magnet using a standard low-frequency lock-in technique; see methods section and SI section 1. The twist angles ($\theta$) of the devices are estimated by measuring the moiré superlattice carrier density $n_s = 8\theta^2/\sqrt{3}\ a^2$ (a=0.23nm) and are found to be 1.38°, 1.41°, and 1.44° for A, B, and C, respectively; see SI section 1. The 4-terminal longitudinal ($R_{xx}$) and transverse ($R_{xy}$) resistance as a function of carrier density and magnetic field at zero displacement field, $D$, for devices A, B, and C, are shown in extended Fig 1. The $R_{xx}$ ($R_{xy}$) for devices A, B, and C are measured across the contacts configuration shown in Fig. 1b; see SI section 1 for more information. Similar to previous reports on TTG, we observe Chern Insulator (CI) states, giving rise to peaks of resistance screened by the Dirac spectrum of MLG at a moiré filling factor, $\nu$, of $\nu = 1, 2, 3, 4, -2, -4$ in all devices. Robust superconductivity pockets on both electron and hole-doped regions, with a maximum critical temperature of $T_{SC} \sim 1.3$ K (electron-doped at v~2.6 and D~0.38V/nm) are observed in device C and across the Josephson junction of devices B and C (see extended Fig 3 and supplementary Fig. S6(a)).

By examining the variation of $R_{xy}$ as a function of the out-of-plane magnetic field, $B_z$, we mapped out the linear Hall dependence as a function of the filling factor. However, unexpectedly, at certain filling factors near the charge neutrality point (CNP), a sudden increase in the slope develops (at $B_z$=0), as highlighted by the purple line in Fig. 1c, measured at $\nu = -0.45$ and $D = 0\ V/nm$. As the filling factor or displacement field increased, the slope reverts to its expected linear behavior. To further explore the phase space where the increased slopes in $R_{xy}$ occur as a function of the filling factor and displacement field, we measured $R_{xy}$ as a function $B_z$ and $\nu$ at $D = 0\ V/nm$ and $D = 0.6\ V/nm$. The Landau-level fan diagrams of $\frac{dR_{xy}}{dB_z}$ for device A, as a function of $\nu$ at $D = 0\ V/nm$ and $D = 0.6\ V/nm$ are shown in Fig. 1d and 1e, respectively. These plots of $\frac{dR_{xy}}{dB_z}$ highlight the regions with increased slopes by the darker color regions around $B_z$=0. In Fig. 1d, as $\nu$ increases away from CNP, the diminishing intensity of the darker regions becomes evident, indicating the suppression of the jumps in $R_{xy}$. At high displacement field $D = 0.6\ V/nm$ the $R_{xy}$ jumps in the vicinity of CNP likewise decrease, as seen in Fig. 1e; see also the blue line cut at $\nu = -0.45$ in Fig. 1c. The resistive peaks at $\nu = 1, 3, -3$ are also no longer observed. Finally, the $\frac{dR_{xy}}{dB_z}$ slope extracted after subtracting the linear Hall effect is shown in the inset of extended Fig. 2 as a function of $\nu$ for all three Hall bar devices. The maximum slope is observed on the hole doping side around $\nu = -0.45$, and the jumps in $R_{xy}$ disappear for $\nu$ >2 and $\nu$<-2, see black line cut in Fig. 1c at $\nu = -3.6$ and $D = 0$ V/nm. We observed a slight hysteresis behavior (see Figure S4 of SI), likely due to the presence of multiple magnetic domains.

In Fig. 1f and 1g, we extract the Hall carrier density, $n_H = 1/e * \frac{dR_{xy}}{dB_z}$, as a function of the filling factor, $\nu$ at $B_z=0$, see SI section 2 and Fig. S2 for more information. Fig. 1f shows $n_H$ vs. $\nu$ plot at $D = 0\ V/nm$, where we observe resets at integer $\nu = 1, \pm 2, \pm 3, \pm 4$, indicative of Chern insulator states at these integer fillings[17,26]. However, instead of a linear slope in $n_H$ vs. $\nu$, an offset is observed from both doping regions of the CNP, emanating from fractional fillings of $\nu \sim -0.5$ and $0.7$, see SI section 1 and Fig. S2(a) for more information. This abrupt change in DOS hints towards an intrinsic effective magnetic field, which might originate from spontaneous symmetry breaking of degenerate bands near the CNP[27,28]. We repeated the analysis at $D = 0.6\ V/nm$, in Fig. 1g, where the expected linear Hall magnetic field dependence was restored around CNP.

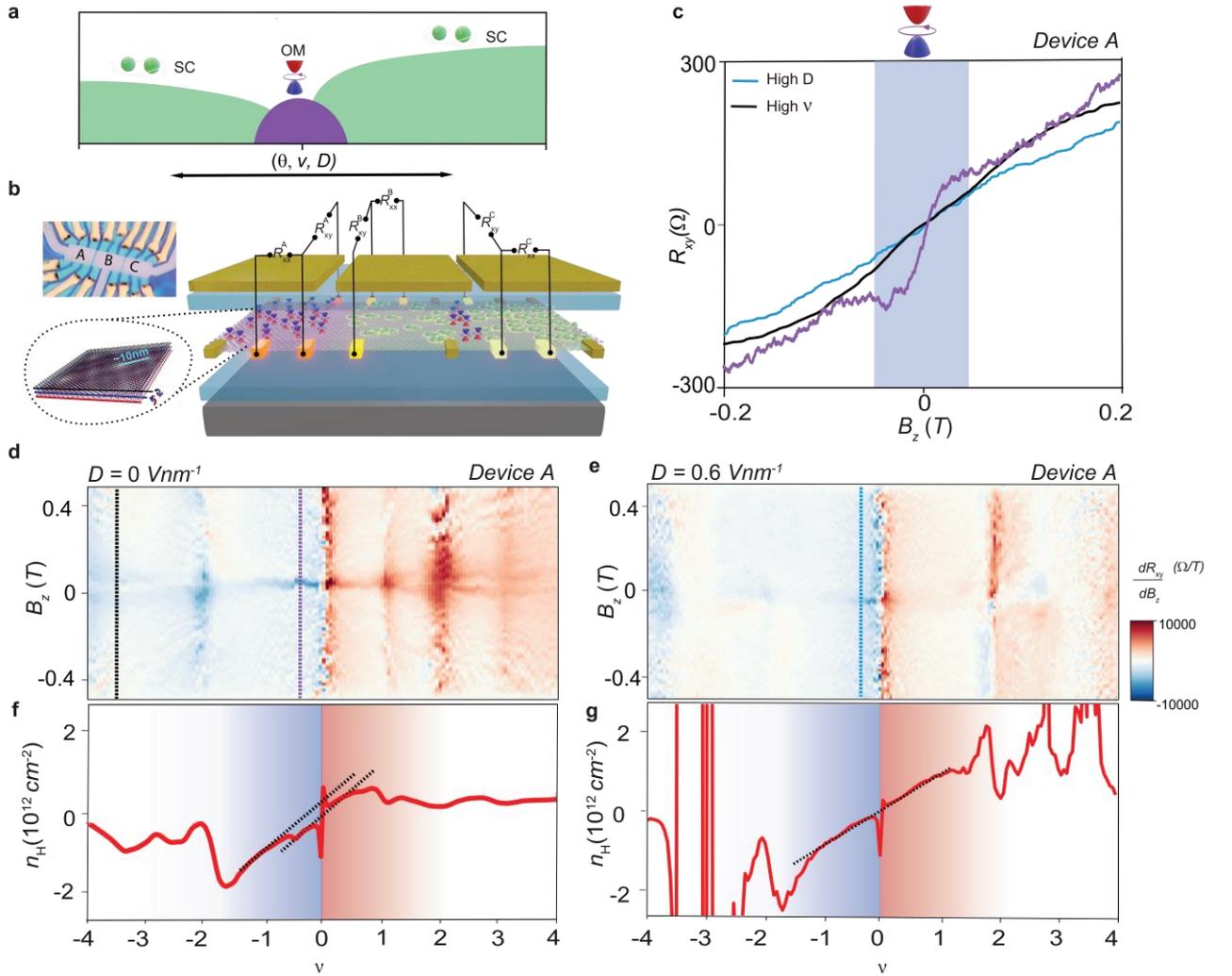

**Figure 1: Density tunable jumps in Hall measurements.** **(a)** Schematic phase space of superconductivity and orbital magnetism in alternating TTG as a function of twist angle θ moiré filling factor $\nu$ and displacement field D. **(b)** Schematic diagram of alternating TTG Hall bar devices and in situ gate defined Josephson junctions between them. The TTG is encapsulated by hBNs on both sides, metal (Au/Ti) gated on top and bottom gated using oxidized silicon (Si/SiO$_2$ 285nm). An enlarged picture of alternating TTG and an optical image of the device are shown alongside. The moiré length is extracted using the formula $\lambda = a/[2 \sin(\theta/2)]$ (a = 0.246 nm and θ=1.40º). **(c)** The line scans of $R_{xy}$ as a function of $B_z$ at three distinct ($\nu$, D) values *i.e.* purple line at $\nu = -0.45$ and $D = 0\ V/nm$, black line at $\nu = -3.6$ and $D = 0$ V/nm (high $\nu$), blue line at $\nu = -0.45$ and $D = 0.6\ V/nm$ (high D) for device A. Corresponding colors of line cuts are shown

in Fig. (d) and (e). **(d)** and **(e)** The Landau fan diagram of $B_z$ v/s $\nu$ for differential Hall resistance w.r.t magnetic field ($\frac{dR_{xy}}{dB_z}$) at D=0V/nm and D=0.6 V/nm respectively for device A. **(f)** and **(g)** The variation of Hall carrier density ($n_H$) with $\nu$ at $B_z$=0 at D=0V/nm and D=0.6 V/nm respectively for device A.

The mirror-symmetric configuration of graphene layers in TTG enables the exploration of hybridization effects between the Dirac and flat band sectors using a displacement field. Figures 2a and 2b present the phase space of $\frac{dR_{xy}}{dB_z}$ as a function of $B_z$ and $D$, at $\nu = -0.45$ and $0.7$, respectively for device A. Darker colors correspond to larger magnitudes of $|\frac{dR_{xy}}{dB_z}|$. In Fig. 2a, at $\nu = -0.45$, a jump in $R_{xy}$ is observed at $D = 0\, V/nm$ (indicated by the brown color), as also seen in Fig. 1c. This jump is suppressed as the displacement field increases beyond $|D| > 0.5\, V/nm$. In Fig. 2b, at $\nu = 0.7$, an opposite slope in $R_{xy}$ is observed, with jumps in $R_{xy}$ (indicated by the blue color) being suppressed for $|D| > 0.25\, V/nm$. These findings suggest that the displacement field in TTG enables a tunable hybridization between the monolayer Dirac sector and the flat bands sector, leading to the vanishing of OM.

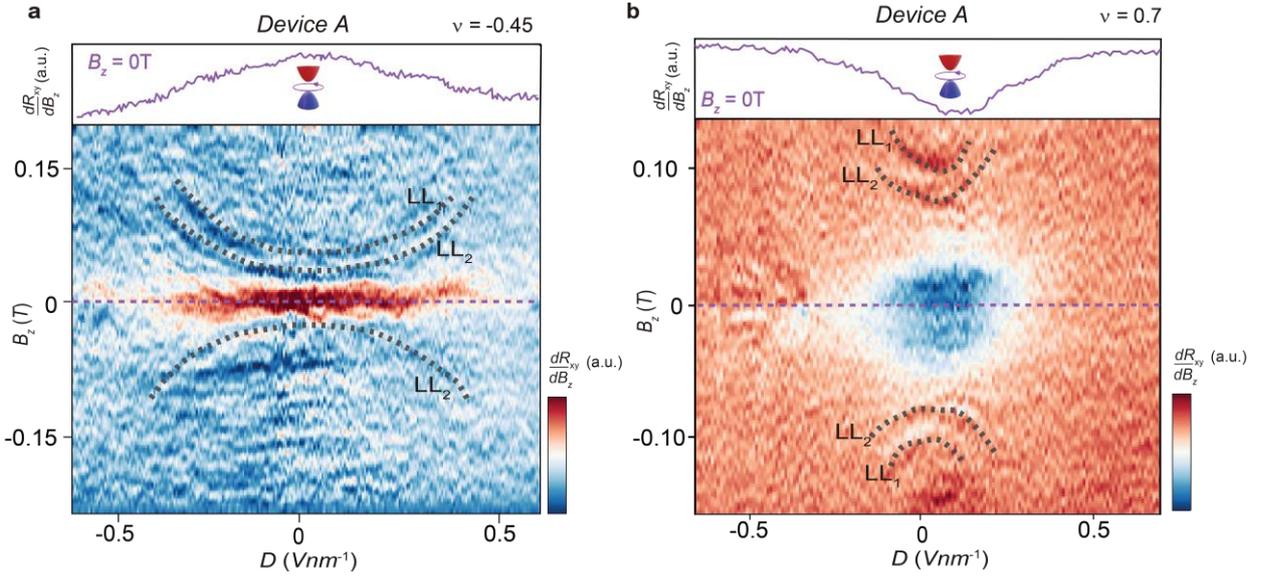

**Figure 2: Displacement field tunable jumps in Hall measurements.** The displacement field (D) dependence of $\frac{dR_{xy}}{dB_z}$ while sweeping $B_z$ around 0T for device A. The moiré filling factor ($\nu$) is fixed at **(a)** $\nu$=-0.45 and **(b)** $\nu$=0.70. The dark brown (Fig. a) and blue regions (Fig. b) nearby $B_z = 0T$ show the maximum change in $\frac{dR_{xy}}{dB_z}$, which disappear at high D. Insets show the line cuts at $B_z = 0T$, corresponding to purple dash lines in figures. The dotted curved black lines represent the variation of Landau levels 1 ($LL_1$) and 2 ($LL_2$) of the monolayer Dirac sector with D.

Band mixing is thought to play a crucial role in determining the nature of the correlated low-energy phases[29,30]. To quantify the hybridization of the two sectors, we analyze the parabolic dashed curves highlighted in Fig. 2. These curves correspond to the Shubnikov-de Haas (SdH) oscillations of the MLG Landau levels ($LL_s$), at specific Fermi energies $E_F$ for a constant $\nu$ [6,30,31]. By examining the shift in the MLG $LL_s$, we can infer the degree of hybridization between the two sectors. If we assume the Fermi velocity, $v_F = dE/dk$ of MLG Dirac cone is $10^6$m/s at D=0V/nm, we can estimate the percentage change in $v_F$ with increase in D, see SI section 2 for more details. At $\nu = -0.1$, $-0.45$ and $-0.70$ for an increase in D from 0 to 0.20 $V/nm$, $v_F$ decreases by ~30%, 20% and 12% respectively. The MLG LL curvature lines disappear at D~0.45, 0.4 and 0.25 V/nm at $\nu = -0.1$, $-0.45$ and $-0.70$ respectively, coinciding with the disappearance of OM, See Fig. 2a, 2b and S3a. Thus, we conclude nearby CNP, a higher value of D is required for the hybridization of MLG with the flat bands and OM persists over longer ranges of D. However, as we increase the density, a smaller value of D is sufficient for hybridization since the Fermi energy is closer to the overlap between MLG and flat bands[31]. In ref [29] singlet superconductivity is predicted to emerge near CNP at low displacement fields in TTG. The observation of OM near CNP in our samples might explain the absence of such a singlet superconducting state due to competing ordering tendencies.

Magnetic ordering is a result of exchange interactions due to time-reversal symmetry breaking and can be observed due to spin magnetic moments or electron circular motion-driven orbital magnetic moments. While spin magnetic moments are typically isotropic, orbital moments are highly anisotropic. Previous studies have indicated that magnetism in graphene-based heterostructures is predominantly of orbital origin [19,21–24,32]. To determine the dominant mechanism behind the magnetic state, we measured $R_{xy}$ while varying the angle between the magnetic field and the sample plane. Fig. 3a shows $R_{xy}$ as a function of $\vec{B} = B\cos(\theta) \cdot \hat{z} + B\sin(\theta) \cdot \hat{x}$ at $\nu = -0.45$ and $D = 0 V/nm$ for device A. The angle $\theta$ between the sample and $\vec{B}$ is varied from 90° to 0° in the steps of 15°. The steepest slope in $R_{xy}$ is observed when $\vec{B} = B_z \cdot \hat{z}$, while no jump is detected when $\vec{B} = B_x \cdot \hat{x}$, indicating the highly anisotropic nature of the magnetic state. See fig. S5 of SI for the same measurements on device C and D. The inset of Fig. 3a shows $R_{xy}$ measured as a function of $B_z$ while a fixed $B_x$ =0,0.5 and 1T is applied. The amplitude of the $R_{xy}$ jump remains unchanged even with $B_x$ applied up to 1T. This high out-of-plane anisotropy of the magnetic moments makes an orbital origin highly likely.

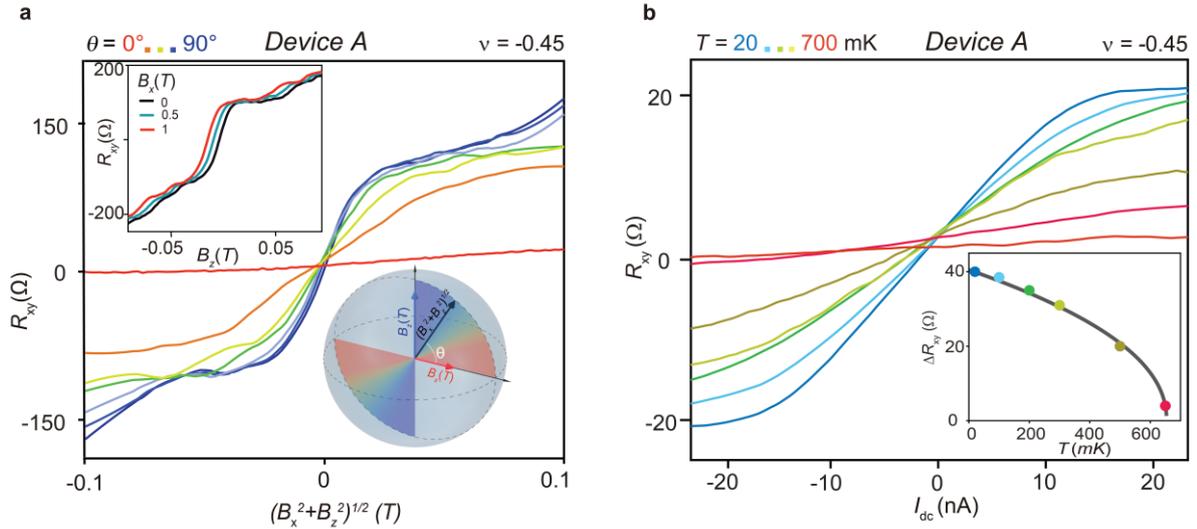

**Figure 3: Confirming the orbital nature of the jumps in $R_{xy}$.** (a) The $R_{xy}$ vs $B = (B_x^2 + B_z^2)^{1/2}$ at varying $\theta$ between sample and B from 90° (blue) to 0°(red) in steps of 15° for device A. Inset shows the jumps in $R_{xy}$ as a function of $B_z$ at constant in-plane field $B_x$ =0,0.5 and 1T. (b) The temperature dependence of jumps in $R_{xy}$ as a function of DC bias current ($I_{dc}$) at temperatures ranging from 20mk (blue) to 700mK (red) measured at $B_z = 0T$ for device A. Inset shows a fit of the jump height in $R_{xy}$ ($\Delta R_{xy}$) to the Curie Bloch equation: $\Delta R_{xy} = p(1 - T/T_{OM1})^\alpha$.

In studies on TBG aligned with hBN, orbital magnetic moments were polarized using an external DC bias ($I_{dc}$) superimposed on a small AC signal[21,23]. We measured $R_{xy}$ at $\nu = -0.45$ and $D = 0V/nm$ with an AC excitation of approximately 1nA while varying the $I_{dc}$ from -25nA to +25nA at $B_z = 0T$. A jump in $R_{xy}$ is observed at 20mK, and the amplitude of this jump decreases with increasing temperature. Fig. 3b shows the measurements taken up to 700 mK, with no jump observed above approximately 650 mK. The difference in $R_{xy}$ at $I_{dc}$ =-20nA and +20nA *i.e* $\Delta R_{xy}$ can be fitted to the Curie Bloch equation $\Delta R_{xy} = p(1 - T/T_{OM1})^\alpha$ (here $p$ is a proportionality constant, $\alpha$ is the fitting exponent and $T_{OM1}$ is the critical temperature of the magnetic state) yielding $T_{OM1}$=654±6 mk, $p$=40.5±1Ω and $\alpha = 0.45\pm 0.05$.

The observed jumps in Hall resistivity as a function of bias current indicate the coupling between electric fields generated by currents in the conducting bulk and the magnetization of sample [33]. Recent observations of $I_{dc}$ induced switching of $R_{xy}$ in TBG aligned to hBN are attributed to an interplay between extrinsic breaking of sublattice symmetry (due to alignment with hBN), breaking of rotation symmetries (due to strain), and intrinsic spontaneous time-reversal symmetry breaking [33]. Rotation symmetry ($C_3$) breaking in TTG was recently reported *via* observation of transport non-reciprocity[34] and a nematic semimetal ground state at CNP[31]. Our results implicate the orbital magnetic moments, due to the spontaneous breaking of valley isospin symmetry near the CNP, as the culprits in facilitating the significant current-magnetization coupling. Indeed, TTG has shown evidence of exchange interactions for $v < 1.5$, while at higher densities, the charging self-energy becomes dominant[31].

We now turn to investigate the impact of OM on supercurrent flow by incorporating it as the weak link in an electrostatically defined Josephson junction (JJ)[35,36], leveraging the tunable superconductivity in TTG. Valley polarization-driven OM resulted in asymmetric Fraunhofer patterns in previous graphene-based twisted devices [19,32]. Extended Fig. 3a-c presents the $R_{xx}$ measurements for all three devices under various displacement fields. As observed previously, superconductivity is enhanced at high displacement fields [7,8] (in device C) and suppressed at twist angles between 1.38° and 1.41°, consistent with earlier studies[19,37]. The phase space of $R_{xx}$ due to the JJ weak link between devices B and C (JJ2 contacts 5-6 of JJ2, see supplementary Fig. S1) is shown in extended Fig. 4a. We observe clear resistive states and superconductivity across JJ (yellow dashed lines correspond to left and right-side *v* in extended Fig. 4a). The vertical resistive lines in pink correspond to the filling factor ($v_j$) of the weak link region with only the back gate applied.

Figures 4a, 4b, and 4c show the critical current as a function of out-of-plane magnetic field, $B_z$, for various phases of the weak link. The differential resistance ($dV/dI$) was measured at 20mK with an AC excitation of 1nA while sweeping the DC component of the current, $I_{dc}$. The left and right sides of JJ are tuned to a superconducting state (S) corresponding to *v*~2.6 using a combination of top and bottom gates. The weak link is tuned to a superconducting state ($S'$) $v_j \sim -2.7$ ($D_j$ ~-0.3V/nm), normal metallic state (N) $v_j \sim -1.7$ ($D_j$ ~-0.19V/nm) and orbital magnetic state (OM) $v_j \sim -0.45$ ($D_j$ ~-0.05V/nm) using the back gate only, as shown in the insets of Figs. 4a, 4b and 4c respectively. The line cuts of $dV/dI$ vs $I_{dc}$ at $B_z$=0G for these JJ configurations are shown in supplementary Fig S6 (b).

Fig. 4a presents a typical 2D measurement for the S|S'|S configuration, where no Fraunhofer pattern is observed, confirming the presence of a relatively uniform superconducting region extending from the reservoirs across the weak link. As expected in this configuration, the maximum positive superconductivity critical current ($I^+_{c,max}$) and maximum negative superconductivity critical current ($I^-_{c,max}$) curves are symmetric along $B_z$ axis, see inset of Fig.4a. In the S|N|S configuration, a supercurrent is observed across the JJ, exhibiting Fraunhofer oscillations as a function of $B_z$ (Fig. 4b). The high-field periodicity of these oscillations, for a 2D superconductor, is predicted to follow $\Delta B \sim 1.8\Phi_o/w^2$, where $\Phi_o$ is the flux quantum and $w$ the lateral width of the JJ [38], which for our device evaluates to $\Delta B$~10G. We observe $B_z$-induced oscillations with a width of ~30G for the central maxima and a periodicity of ~10G for the high-field oscillations. The Fraunhofer pattern also shows symmetric $I^+_{c,max}$ and $I^-_{c,max}$, as seen in the inset of Fig. 4b. Remarkably, as we move to the S|OM|S configuration, we observe asymmetric $I^+_{c,max}$ and $I^-_{c,max}$ along the $B_z$ axis, (Fig. 4c). The asymmetry of the

Fraunhofer pattern decreases with an increase in temperature, as shown in the extended Fig. 4b, and a symmetric pattern is obtained at ~650mK.

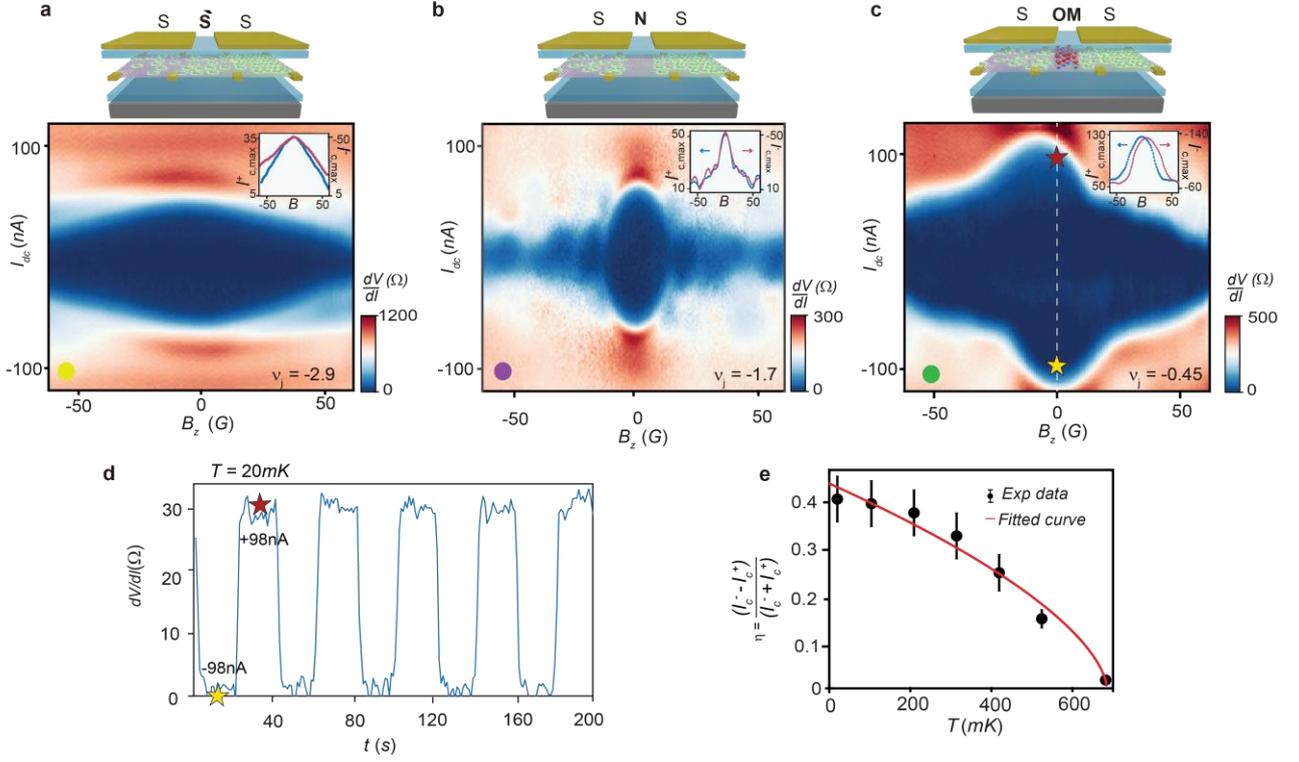

**Figure 4: Gate defined Josephson junction as a probe for orbital magnetism.** The Fraunhofer patterns measurements across JJ2 (contacts 5-6, see SI Fig. S1) , left and right side of JJ tuned to superconducting state (S) at $v\sim2.6$ and weak link tuned to **(a)** $v_j \sim -2.7$ forming S|S'|S JJ, **(b)** $v_j \sim -1.7$ forming S|N|S JJ and **(c)** $v_j \sim -0.45$ forming S|OM|S JJ. Insets on top of the figures show the schematics corresponding to JJ configurations. Top right insets show the line cuts of the positive SC critical current ($I^+_{c,max}$) and negative SC critical current ($I^-_{c,max}$) extracted from corresponding Fraunhofer patterns. **(d)** The $dV/dI$ ($\Omega$) characteristics of S|OM|S JJ measured at $B_z=0$ G while alternating $I_{dc}$ between $I^-_c$ (-98nA) and $-I^-_c$ (+98nA) after every 20 seconds at 20mK. **(e)** The measure of asymmetry in critical currents $\eta = \frac{(I^-_c - I^+_c)}{(I^-_c + I^+_c)}$ is plotted as a function of temperature (black data points) at $B_z=0$ G. The red line corresponds to fit the Curie Bloch equation $\eta = k(1 - T/T_{OM2})^\beta$.

We switch from the superconducting (SC) to the normal state at $B_z=0$, by alternating $I_{dc}$ from 98nA to -98nA, respectively, illustrated by white dashed line in Fig. 4c. Fig. 4d shows the $dV/dI$ characteristics, at 20mK, plotted over time. The resistance across the JJ switched between 0 $\Omega$ (SC state) and 29±4 $\Omega$ (normal) following the SC diode effect, disappearing at 650mK (see SI Fig. S7a). In Fig. 4e, we quantify the asymmetry of the Fraunhofer pattern by calculating $\eta = \frac{(I^-_c - I^+_c)}{(I^-_c + I^+_c)}$, where $I^-_c$ and $I^+_c$ are the critical currents measured across for various temperatures (see SI Fig. S6c). As the temperature increases, $\eta$ gradually decreases, reaching zero at 650mk, where the Fraunhofer pattern becomes symmetric, see extended Figs. 4(b-c) for additional details. By fitting $\eta$ to the Curie Bloch equation, $\eta = k(1 - T/T_{OM2})^\beta$, where $k$ is a proportionality constant, $T$ is the temperature, and $T_{OM2}$ the critical temperature, , we obtained the following parameters: $k$=0.44±0.06, $\beta$~0.6±0.2, and $T_{OM2}$~650±20mK.

The estimated values of $T_{OM2}$ matches well with the temperature estimated from $R_{xy}$ vs. $I_{dc}$ fits in Fig. 3b. This indicates that the asymmetry in the Fraunhofer pattern observed in the S|OM|S configuration is related to orbital

magnetism and possibly magnetic domain boundaries motion. We also note that other configurations, Figs. a-b, are symmetric, ruling out the possibility of an intrinsic device-related phenomenon[32,39].

We observed OM in TTG devices of intermediate twist angles of 1.38° and 1.41°, in the vicinity of CNP, without any clear signs of superconductivity in these devices [19,37]. In contrast, a third device with a twist angle of 1.44° - closer to the magic angle – exhibits superconductivity and OM. We used two independent probes to detect OM: (1) Hall effect measurements and (2) Josephson junctions. The OM signatures obtained from these methods are consistent with the Curie-Bloch equation and follow a power law that reveals a common magnetic ordering temperature of approximately 650 mK. Our sample uniquely exhibits a smaller energy scale for the symmetry broken state ($T_{OM2}$~650mK) than that of the superconducting state ($T_{SC}$~1.3K), see extended Fig. 5a and S6a. The JJ is dissipative at low temperatures and upon heating above $T_{OM2}$, resistance drops by orders of magnitude, and superconductivity prevails over OM. This indicates that the mechanism for spontaneous valley polarization-driven OM observed in our samples is likely different from the anomalous Hall effect observed at integer filling factors in twisted graphene systems[21,23–25,34].

We mapped out the phase space of OM as a function of the filling factor (ν) and displacement field (D). Extended Fig. 5a shows the general phase diagram of OM and SC as a function of *ν* and displacement field D. The OM is most prominent near the CNP and decreases as we approach ν = ±2, where superconductivity emerges. As D increases, OM weakens while superconductivity becomes more robust, revealing a complementary relationship in their phase spaces. Using the Landau levels (LLs) of monolayer graphene (MLG) as a probe, we also demonstrated the variation in hybridization strength between the flat bands and Dirac cones as a function of D. Extended Fig. 5b shows the distribution of electron-doped superconductivity (e-SC), critical temperature ($T_{SC}^e$) and OM ordering temperature ($T_{OM}$) data points as a function of twist angle θ for alternating TTG devices taken from literature and this work. The green and purple colored phase space represents the e-SC and OM, respectively. The competition between magnetic and superconducting phases varies with the twist angle, resulting in a range around 1.4°, where superconductivity is suppressed at all densities while clear OM signatures persist. To further explore OM, we measured two additional TTG devices with twist angles of approximately 1.3° and 1.5°, which show no OM but exhibit superconductivity (see SI Section 3 and Fig. S9, S10, and S11 for further details).

In summary, our measurements reveal a rich phase diagram where both superconductivity and magnetic ordering appear, governed by carrier density, displacement field, and twist angle in TTG. These findings indicate that alternating TTG has a rather dissimilar phase diagram compared to TBG. The D-driven change in hybridization strength of flat bands with Dirac cone changes the ground state and dictates the stabilized correlated phase. The complementary phase space of OM to superconductivity might help explain the origin of superconductivity in TTG heterostructures. Particularly surprising is our finding that the usual sequence of energy scales is reversed in the TTG phase diagram, with superconductivity onsetting at higher temperatures than OM. Such behavior could, for example, result from critical triplet pairing fluctuations at the boundary of the superconductivity phase, giving rise to ferromagnetic correlations, which would constitute an intriguing role reversal compared to the typical phenomenology observed for competing instabilities. From a general physics perspective, the coexistence of magnetism and superconductivity within a single material platform is highly significant. This interplay opens exciting possibilities for realizing exotic quasiparticles in van der Waals-based hybrid devices, potentially advancing the search for non-abelian anyons and novel topological states.

## Methods

### Stacking and Device fabrication

The TTG stack is prepared using dry-transfer method. The hBN and graphene flakes are exfoliated on clean Si/SiO$_2$ (285 nm) substrate. The number of graphene layers are determined by examining the FWHM of 2D peak Raman spectra (WITec alpha300 R) using 532nm laser. Monolayer graphene ~100 X30µm is cut into three pieces separated by ~5 µm gap using 1064nm Raman laser. The clean hBN crystals are examined using an optical microscope and dark field microscopy. The crystallographic axes of hBNs are determined using straight edges. Stamps for picking the flakes are prepared by placing polycarbonate (PC) thin films on polydimethylsiloxane (PDMS) dome stamps. The top hBN flake (~25nm) is picked at 100°C and the graphene flakes are picked up at 40°C. The transfer stage holding the Si chip with vacuum is rotated to ~1.45° and ~-1.45° to obtain the mirror-symmetric configuration of TTG. The bottom hBN (~30nm) is picked up at 50°C and the final stack dropped on clean Si/SiO$_2$ (285 nm) substrate at 180 °C. The melted PC on the stack is cleaned using chloroform and stack annealed in a vacuum at 350 °C to move the air bubbles, release strain, and remove impurities on top of the stack. The stack's contact mode cleaning and hBNs' thicknesses are determined using Bruker atomic force microscope. Jeol JBX9300-FS e-beam lithography is used to define metal top gates and JJs of lateral width~2µm. The Cr(4nm)/Au(16nm) metals are deposited with an e-gun evaporator for top gates. The edge contacts to graphene are made using CHF$_3$/O$_2$ plasma in RIE and consecutively depositing the Cr(2nm)/Au(65nm) metals in the angle rotator e-gun evaporator. The device is etched into Hall bar geometry using CHF$_3$/O$_2$ plasma in RIE.

### Measurements

Bluefors LD400 dilution refrigerator with RC and RF filtering having base temeperature~10mk is used to measure the transport characteristics. The Q-devil sample puck with additional filtering is used to mount the sample to the fridge. The four-probe measurements are performed using standard lock-in techniques using $I_{ac}$ ~1nA rms (100MΩ resistor) and 11.377Hz frequency. The Femto voltage amplifiers are used at room temperature to amplify signals from fridge to SRS830 and SRS865A lock in amplifiers. Keithley 2400 source meters are used to apply top and bottom gate voltages. The bottom gate volage ($V_{bg}$) and top gate voltage ($V_{tg}$) are converted to ($n$) and ($D$) using electrostatic equations $n = \frac{\varepsilon_t \varepsilon_0 (V_{tg} - V_{to})}{e d_t} + \frac{\varepsilon_b \varepsilon_0 (V_{bg} - V_{bo})}{e d_b}$ and $D = \frac{\varepsilon_t \varepsilon_0 (V_{tg} - V_{to})}{d_t} - \frac{\varepsilon_b \varepsilon_0 (V_{bg} - V_{bo})}{d_b}$ ($\varepsilon_b$, $\varepsilon_t$ dielectric constant of bottom and top hBN ~ 3.6 ; $\varepsilon_0$: permittivity of air; $e$ : charge of electron; $d_b$, $d_t$ : thickness of bottom (30nm) and top hBN (25nm); $V_{bo}, V_{to}$ are the bottom and top gate voltages of charge neutrality point at zero magnetic field). From the Landau fan diagram of $R_{xx}$ and $R_{yx}$ we find the moiré superlattice carrier density $n_s = 4.65 \times 10^{12} \text{cm}^{-2}$, $4.72 \times 10^{12} \text{cm}^{-2}$ and $4.80 \times 10^{12} \text{cm}^{-2}$ corresponding to the twist angle of $1.38°$, $1.41°$ and $1.44°$ using equation $n_s = 8\theta^2/\sqrt{3} \, a^2$ (a=0.234nm), see supplementary section 1 for more information. AMI 9-1-1 vector magnet is used to study magnetic field direction dependence on $R_{yx}$. For $dV/dI$ measurements AC excitation voltage of ~0.1V rms is applied using lock in amplifier and dc bias voltages is applied using Yokogawa voltage source meter through 100MΩ resistor.

**Analysis of Hall data and fittings:** We calculated Hall carrier density using equation, $n_H = 1/e * \frac{dR_{xy}}{dB_z}$, (here, $e$ is elementary charge), the $\frac{dR_{xy}}{dB_z}$ is calculated at low $B_z$ range : -0.1T $\leq B_z \leq$ 0.1T. The high $B_z$ range: -0.5T $\leq B_z \leq$ 0.5T fitted data is shown in SI Fig. S2. The amplitude of $R_{xy}$ slope w.r.t $B_z$ shown in extended Fig 2 is calculated by subtracting linear slope of $R_{xy}$ vs $B_z$ from $\frac{dR_{xy}}{dB_z}$. To quantify the change in the $v_F$ with displacement field, we estimate Fermi energy $E_F$ by employing a single-particle equation for the MLG $LL_s$ spectrum, $E_F = sgn(L_n)v_F\sqrt{(2e\hbar |L_n| \times B)}$; where $L_n$ is Landau level index, $v_F$ is Fermi velocity of the monolayer Dirac cone, $e$ is the elementary charge, and $\hbar$ is reduced Planck's constant. First, we calculated $E_F$ for $LL_1$ at D=0V/nm as ~-8meV at $v$ ~-0.45, ~-4meV at $v$ ~-0.10 and ~11meV at $v$~0.70 by using $v_F = 10^6$m/s, see SI Fig S3 for more details. Since $E_F$ is constant for given $LL$, the change in curvature is used to estimate $v_F$ change with the displacement field. The $R_{xy}$ jump amplitude as a function of dc bias current is fitted to the Curie Bloch equation using non-linear least square fitting method.

**Josephson junction analysis:** The phase space of JJ as a function of the top gate and the back gate is shown in extended Fig 4. The carrier density ($n_j$) and displacement field ($D_j$) across JJ is calculated using electrostatic equations $n_j = \frac{\varepsilon_b \varepsilon_0 (V_{bg} - V_{bo})}{e d_b}$ and $D_j = \frac{\varepsilon_b \varepsilon_0 (V_{bg} - V_{bo})}{d_b}$ ($\varepsilon_b$ dielectric constant of bottom ~ 3.6 ; $\varepsilon_0$: permittivity of air; $e$ : charge of electron; $d_b$: thickness of bottom (30nm) ; $V_{bo}$ are the bottom gate voltages of charge neutrality point at zero magnetic field). The moiré filling factor $v_j$ of JJ is calculated using the equation $v_j = n_j / n_s$ corresponding to $n_s = 4.80 \times 10^{12}$cm$^{-2}$. The diode effect at $B_z = 0T$ is measured using 1nA ac excitation and switching dc bias current between superconducting critical current $I_C$ ($dV/dI$=0 Ω) and -$I_C$. The Curie Bloch equation is fitted to the $\eta$ and resistance using the non-linear least square fitting method.

**Data and materials availability:**

The data supporting the findings in this paper are available from the corresponding author on reasonable request.

**Acknowledgements:** We thank Eli Zeldov and Patrick Ledwith for fruitful discussions. V.B. acknowledges support from the Dean of the Faculty. Y.R. acknowledges the support from the Quantum Science and Technology Program 2021, the Schwartz Reisman Collaborative Science Program, supported by the Gerald Schwartz and Heather Reisman Foundation, the Minerva Foundation with funding from the Federal German Ministry for Education and Research, and the European Research Council Starting Investigator Grant Anyons 101163917. This work is supported by a research grant from the Goldfield Family Charitable Trust

**Author contributions:** V.B. stacked and fabricated the device. L.R. and Y.R. helped design and improve the device quality. K.W. and T.T. grew the hBN crystals. V.B. L.A. and L.R. set up the fridge and developed the measurement codes. V.B. and L.R. performed the measurements. A.I. prepared the schematic illustrations of the devices. V.B., L.R., L.A., M.B., G.S., Y.O., T.H., and Y.R. contributed to the analysis of the results. V.B., L.R., and Y.R. wrote the paper with input from all coauthors. Y.R. supervised the overall work done on the project

**Competing interests**: The authors declare no competing interests.

**Extended Figures:**

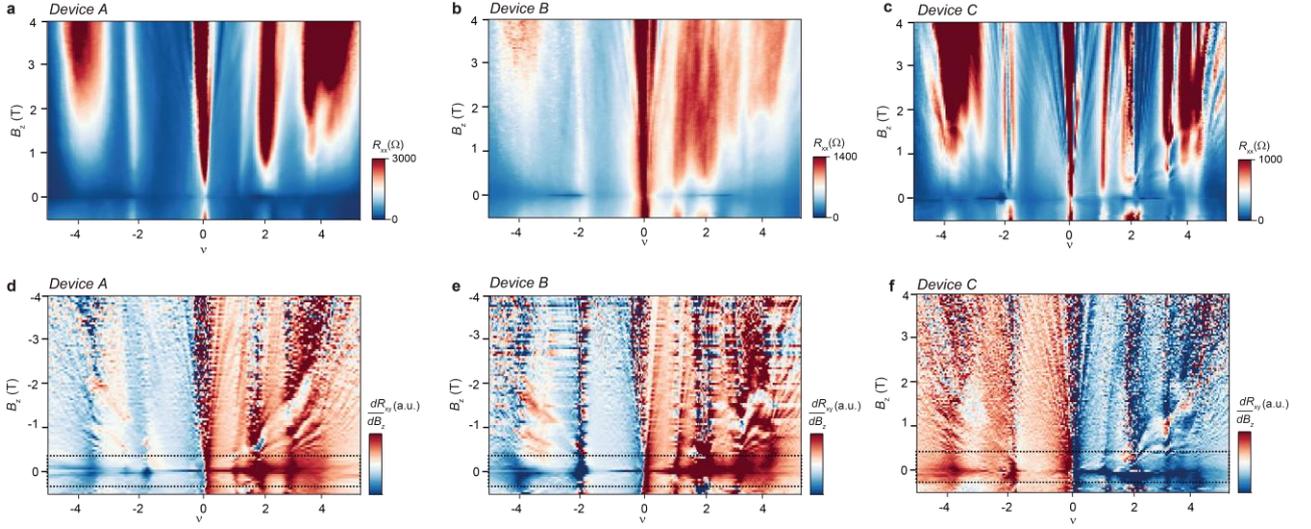

**Extended figure 1: The Landau fan diagrams of $R_{xx}$ and $R_{xy}$.** Measurements are taken at 15mK and D=0V/nm for device **(a)** A (contacts 2-3) **(b)** B (contacts 14-13) and **(c)** C (contacts 6-7). The Landau fan diagram of $\frac{dR_{xy}}{dB_z}$ measured at 15mK and D=0V/nm for device **(a)** A (contacts 15-3) **(b)** B (contacts 14-4) and **(c)** C (contacts 11-6). Dashed black boxes show the slope change in $\frac{dR_{xy}}{dB_z}$ around $B_z$ =0T.

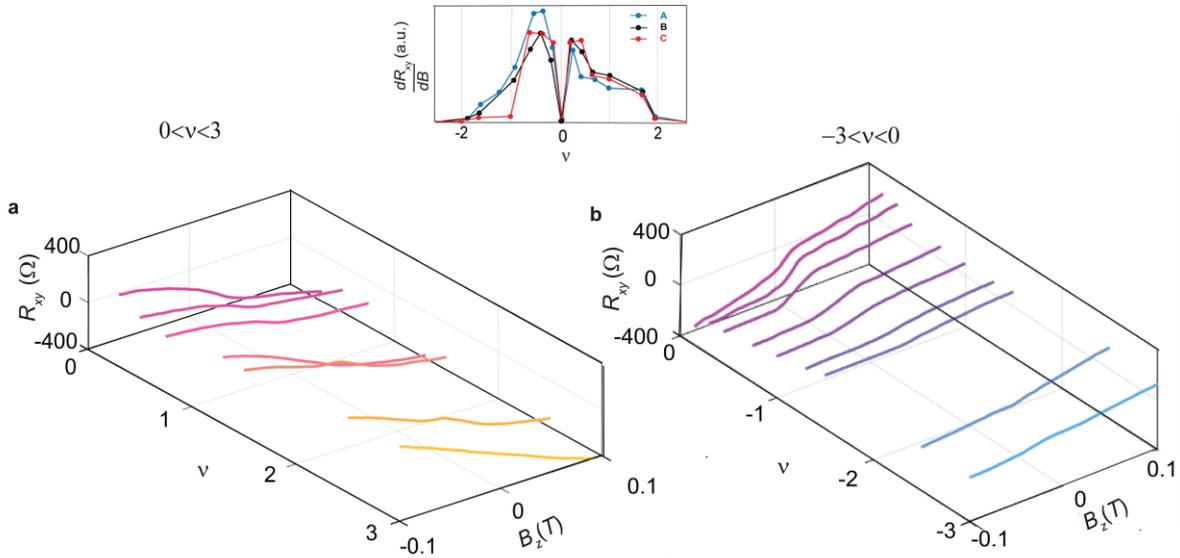

**Extended figure 2: Evolution of jumps in $R_{xy}$ as a function of ν for electron and hole-doped regions.** Line cuts of $R_{xy}$ vs $B_z$ for **(a)** electron and **(b)** hole doping at D=0V/nm for device A. Inset shows the variation of $\frac{dR_{xy}}{dB_z}$ amplitude around $B_z$=0 extracted after subtracting the linear slope of $R_{xy}$ at high $B_z$ as a function of filling factor ν for all three devices. Maxima is obtained in vicinity to CNP on hole doping side.

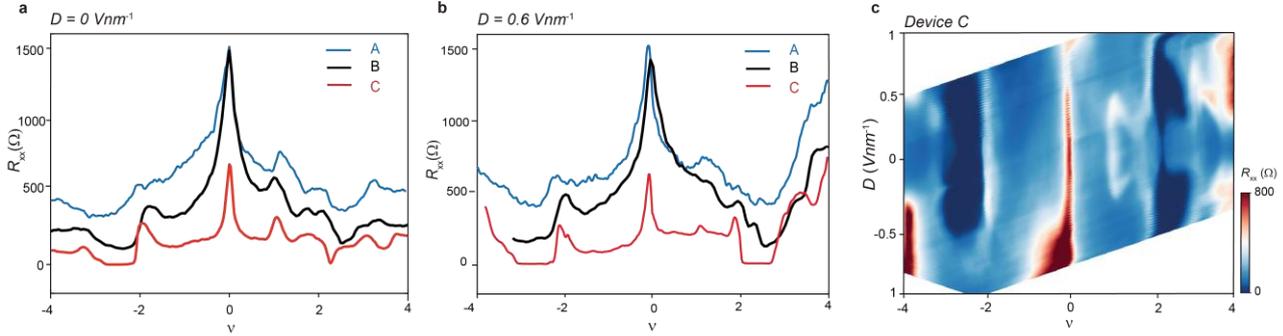

**Extended figure 3: Evolution of superconductivity with twist angle.** Line cuts of $R_{xx}$ vs $v$ for all three devices at T=20mk and D = **(a)** 0V/nm and **(b)** 0.6V/nm. **(c)** D vs $v$ phase space of $R_{xx}$ for device C, at T=20mK.

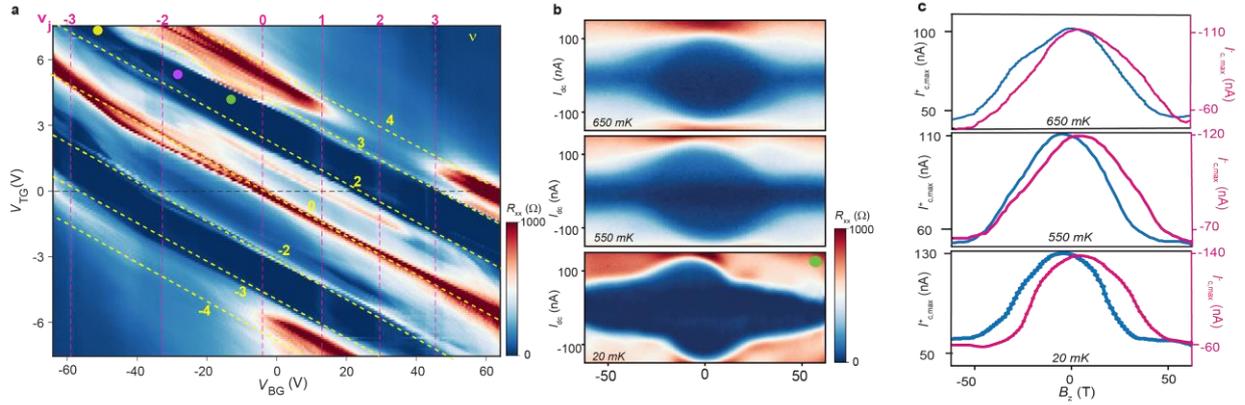

**Extended figure 4: Phase space of Josephson junction and temperature dependence of Fraunhofer asymmetry. (a)** Phase space of $R_{xx}$ for top gate ($V_{TG}$) vs back gate voltage ($V_{BG}$) across the JJ2 at 20mK. The slanted dotted yellow lines correspond to the filling factor ($v$) of regions with both top and back gate *i.e* left and right sides of the JJ. The vertical pink dotted lines correspond to the filling factor of the weak link ($v_j$) region with no top gate. **(b)** The Fraunhofer pattern measured at $v_j$~-0.45 forming S|OM|S JJ at 20mK, 500mK and 650mK. **(c)** Line cuts of the maximum critical current extracted from figure b.

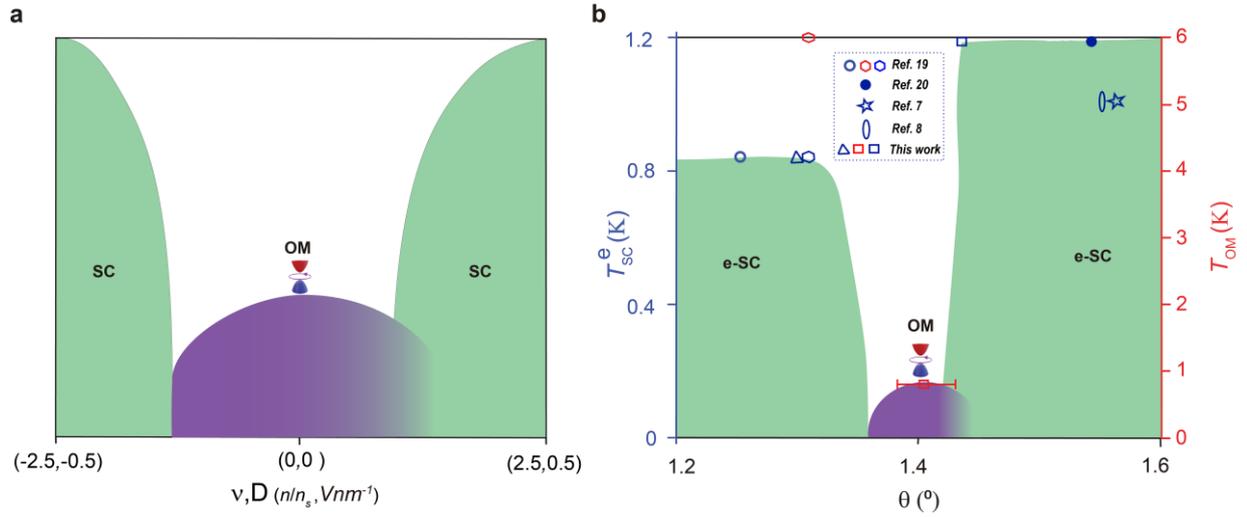

**Extended figure 5: Phase diagrams of superconductivity and orbital magnetism as a function of twist angle, density and displacement filed.** (**a**) General phase diagram of SC and OM with $v$ and displacement field D. The OM is stronger near D ($v$)=0, and gets weaker as we move away, whereas SC is weaker (absent) near D($v$)=0 and gets stronger as we move away. (**b**) Phase space of critical temperature of electron doped superconductivity (e-SC) $T_{SC}^{e}$ (blue data points, corresponding to left y axis) and OM ordering temperature $T_{OM}$ (red data points corresponding to right Y axis) as a function of twist angle θ for alternating TTG devices. The OM peaks around 1.40° and is flanked by e-SC domes on lower and higher angles.